\magnification=1200
\def\msun{{\rm M_{\odot}}}

\def\l{{\bigskip\noindent}}
\def\n{{\noindent}}

\centerline{\bf On `The spectral and temporal properties of an
Ultra--Luminous X--ray source} 
\centerline{\bf in NGC 6946', by Senorita Devi et al.} 
\l 
\l 
\l 
\centerline{A. R. King, Theoretical Astrophysics Group} 

\centerline{University of Leicester, Leicester LE1 7RH, UK}

\l 
{\it Abstract} 
\l 
A recent paper (arXiv:0804.3463) claims that the presence of a soft
X--ray component in an X--ray source in NGC 6946 provides strong
evidence that its black hole mass is definitely $>100\msun$, and more
probably $\sim 400\msun$, even if the source is beamed.

\n I show that the data are instead very compatible with a black hole
mass of only $5\msun$, radiating isotropically (i.e. no
beaming) and a mass transfer rate about 60 times the Eddington value,
i.e. $1.25 \times 10^{-5} \msun\, {\rm yr}^{-1}$. Such a system is
very similar to the probable descendant of the Galactic X--ray binary
Cygnus X--1 once the supergiant companion evolves to fill its Roche
lobe.

\l 
{\it Discussion} 
\l 
The X--ray source NGC 6946 X--7 has an inferred blackbody luminosity
$L_{\rm bb} \simeq 2.8 - 3.7 \times 10^{39}\,{\rm erg~s}^{-1}$ and an
inferred blackbody radius $R_{\rm sph} \simeq 6\times 10^8\, {\rm
cm}$.  Senorita Devi et al. (2008: hereafter SD08) derive a black hole
mass $M = 100 - 400\msun$ by assuming that $R_{\rm sph} = 10GM/c^2$,
corresponding to the inner edge of an accretion disc.

\l
However this is not the only possibility. According to Shakura \&
Sunyaev (1973) an accreting source fed mass at a rate $\dot M$ greater
than the Eddington value $\dot M_{\rm E}$ emits a bolometric
luminosity
$$
L_{\rm bol} \simeq L_E\biggl[1 + \ln\biggl({\dot M\over\dot
M_E}\biggr)\biggr].
$$

\n
Further, the characteristic lengthscale for the blackbody disc
emission is now not a few Schwarzschild radii (as assumed in SD08) but
instead
$$
R = {27\over 4}{\dot M\over \dot M_{\rm E}}
$$ 
(cf Begelman et al., 2006). In addition, Shakura \& Sunyaev (1973)
show that the source expels most of the super--Eddington accretion in
an outflow with toroidal geometry, making geometrical collimation by
some factor $b < 1$ likely (cf King et al., 2001). Thus the high
apparent luminosity of ULXs results from two effects: (a) a
bolometric luminosity raised above $L_{\rm E}$ by the natural logarithm of the
Eddington ratio, and (b) geometrical collimation, i.e.
$$
L_{\rm app} \simeq {L_E\over b}\biggl[1 + \ln\biggl({\dot M\over\dot
M_E}\biggr)\biggr].
$$

\n 
In a recent paper (King, 2008) I showed how one could decide which of
these was the more important effect in those ULXs where a soft blackbody
component is detected, as in NGC 6946 X--7. Consistency between the
observed luminosity $L$ and the inferred blackbody radius (also
affected by beaming as $R_{\rm sph} = b^{1/2}R$) leads to the relation
$$
b = {0.016m_1\over L_{40}}\biggl[1 + \ln\biggl({490R_9\over
m_1}b^{1/2}\biggr)\biggr]
$$ 
For an assumed accretor mass $M_1 = m_1\msun$, inferred luminosity
$10^{40}L_{40}{\rm erg\,s}^{-1}$, and black body radius $10^9R_9\,{\rm
cm}$, this equation can be solved to find the beaming factor
$b$ required to produce the observed apparent luminosity.

\l
In King (2008) I showed that it is generally the Eddington logarithm
which dominates the ULX effect, and beaming is rather mild. With
typical ULX values $m_1 = 10, L_{40} = R_9 = 1$ one finds beaming
factors $b \sim 0.76$ and Eddington ratios $\sim 40$, corresponding to
mass transfer rates $\dot M = 1.0\times 10^{-5}\msun\, {\rm
yr}^{-1}$ (for matter of normal composition). 
These parameters are exactly as expected for the thermal--timescale
mass transfer expected when the supergiant companion in a high--mass
X--ray binary fills its Roche lobe after the standard wind--fed
phase. This shows that one naturally gets a population of
stellar--mass ULXs associated with star formation, as observed for
example in the Cartwheel galaxy (cf King 2002).

\l
For the specific case of NGC 6946 X--7 we have $L_{40} \sim 0.28 -
0.37$ and $R_9 \sim 0.6$. The equation for $b$ only has solutions with
$b <1$ provided that the black hole mass is low ($m_1 < 5$). With
$L_{40} = 0.37$ we find $M_1 = 5\msun, b = 1, \dot M = 1.25 \times 10^{-5}
\msun\, {\rm yr}^{-1} = 60\dot M_{\rm E}$.

\l
I conclude that the observations of  NGC 6946 X--7 can be explained by
a stellar--mass ($5\msun$) black hole receiving thermal--timescale
mass transfer in a high--mass X--ray binary. No beaming is required,
nor indeed is a very high black hole mass.

\l
{\it References}
\l
Begelman M.~C., King A.~R., Pringle J.~E., 2006, MNRAS, 370, 399

\n
King A.~R., 2002, MNRAS, 335, L13

\n
King A.~R., 2008, MNRAS, 385, L113

\n
King A.~R., Davies M.~B., Ward M.~J., Fabbiano G., Elvis M., 2001, ApJ, 
552, L109 

\n
Senorita Devi, A., Misra, R., Santhi, K., Singh, K.~Y,
arXiv:0804.3463, ApJ., in press (SD08)

\n
Shakura N.~I., Sunyaev R.~A., 1973, A\&A, 24, 337

\bye